\documentclass[3p,times, twocolumn]{elsarticle}

\usepackage[utf8]{inputenc}
\usepackage{multirow}
\usepackage{tabularx}
\usepackage{epsfig,color}
\biboptions{sort&compress}

\usepackage{amsmath,amssymb}

\usepackage[figuresright]{rotating}

\def\ln{{\operatorname{ln}}}

\def\rme{{\mathrm{e}}}

\def\Eq{eqn}

\def\ln{{\operatorname{ln}}}

\def\rme{{\mathrm{e}}}

\def\kB{k_{\mathrm{B}}}

\def\Eq{Eq.}

\def\Fig{Fig.}

\newcommand{\trm}[1]{{\textrm{#1}}}

\newcolumntype{C}[1]{>{\centering}m{#1}}

\usepackage{ulem}

\begin{document}

\begin{frontmatter}


\title{Nanochannels and nanodroplets in polymer membranes controlling ionic transport}

\author[SI]{Matej Kandu\v{c}\corref{cor1}}
\ead{matej.kanduc@ijs.si}
\author[ES]{Rafael Roa}
\author[KR]{Won Kyu Kim}
\author[DE1,DE2]{Joachim Dzubiella}

\cortext[cor1]{Corresponding author}

\address[SI]{Jo\v{z}ef Stefan Institute, Jamova 39, SI-1000 Ljubljana, Slovenia}

\address[ES]{Departamento de F\'{i}sica Aplicada I, Facultad de Ciencias, Universidad de M\'{a}laga, Campus de Teatinos s/n, E-29071 M\'{a}laga, Spain}
\address[KR]{Korea Institute for Advanced Study, 85 Hoegiro, Seoul 02455, Republic of Korea}
\address[DE1]{Applied Theoretical Physics -- Computational Physics, Physikalisches Institut, Albert-Ludwigs-Universit\"at Freiburg, Hermann-Herder Strasse 3, D-79104 Freiburg, Germany}
\address[DE2]{Research Group for Simulations of Energy Materials, Helmholtz-Zentrum Berlin f\"ur Materialien und Energie, Hahn-Meitner-Platz 1, D-14109 Berlin, Germany}

\begin{abstract}
Polymer materials with low water uptake exhibit a highly heterogeneous interior, characterized by water clusters in the form of nanodroplets and nanochannels.  Here, based on our recent insights from computer simulations, we argue that water cluster structure has large implications for ionic transport and selective permeability in polymer membranes. Importantly, we demonstrate that the two key quantities for transport, the ion diffusion and the solvation free energy inside the polymer, are extremely sensitive to molecular details of the water clusters. In particular, we highlight the significance of water droplet interface potentials and the nature of hopping diffusion through transient water channels. These mechanisms can be harvested and fine-tuned to optimize selectivity in ionic transport in a wide range of applications. 
\end{abstract}
	
\end{frontmatter}

\section{Introduction}
\label{sec:intro}

Many biological and technological soft materials involve fixed and mobile charges, for which long-range electrostatic forces play a major role in their structure and function. For instance, ion channels embedded in cellular membranes enable an incredibly selective and controllable transmembrane transport, vital for signal transduction in the nervous system and other processes of life~\cite{jensen2010principles, yoder2018gating, kato2018structural}. Importantly, selective transport of ions is paramount also in numerous present-day applications with synthetic materials, ranging from filtration, desalination, battery electrolytes, fuel cells,  biomimetic nanochannels, drug delivery, nanocatalysis, and many more~\cite{li2016designing,lu2019tuning, kusoglu2017new, xie2018bacteriorhodopsin, xin2019high, zhu2020bioinspired, xu2020molecular, liu2020neutralization, widstrom2021water, epsztein2020towards}. A common challenge in these applications is to design and manufacture polymeric membranes in some solvent environments to achieve controllable permeability and selectivity (“permselectivity”) in the charge transport for the desired function.


Molecular mass transport in dense polymeric membranes is typically governed by the solution--diffusion mechanism~\cite{wijmans1995solution}: Small ions and molecules first partition into the polymer matrix from bulk solvent and then diffuse in the polymer under external fields or chemical-potential gradients. 
In this process, ions generally move through networks of nanochannels or porous structures of various complex morphologies and topologies. These pores are often filled with high-dielectric solvents, such as water. Whereas it is clear that bulk solvent is an indispensable medium in which charged molecules are solvated and transported, it is not well known how the solvent behaves and influences charge transport inside the dense, low dielectric polymer matrix. 
The poor understanding is mainly due to the lack of experimental techniques with the necessary temporal and spatial resolution to probe the kinetics of ions. 

Fortunately, computer simulations are a powerful complementary tool that offers insight into those processes~\cite{epsztein2020towards}.
For instance, recent computer simulations revealed that water distributes very heterogeneously in dense polymers in fractal-like cluster structures embedded in the nanometer-sized voids of the polymer matrix~\cite{kanduc2018diffusion, mabuchi2018relationship}. The nanoclustered water was found to act as an important player in the penetrant diffusion and also to govern ion partitioning and permeability~\cite{kanduc2019aqueous, kanduc2021shape, widstrom2021water}. 
Whereas these simulations provide the first unprecedented views on the cluster structure inside polymer matrices and its far-reaching effects on transport, many quantitative details still remain elusive.  For example, how does cluster shape, interfaces, or connectivity affect ion partitioning or transport in detail? How is the water cluster structure affected and controlled by temperature or water volume fraction?  What role does the ‘chemistry’ of ions play, that is, ionic shape, polarity, and charge structure? Are some of these aspects universal and addressable by relatively non-specific continuum concepts, even by empirical laws? 

In this paper, we present our view on the role of aqueous nanoclusters (droplets and channels) in polymer networks on ion diffusion, solvation, and permeability and address a few of the above open questions. The perspective is based foremost on our recent, extensive molecular simulations of ion transport in aqueous poly($N$-isopropylacrylamide) (PNIPAM) systems~\cite{kanduc2018diffusion, kanduc2019free, kanduc2019aqueous, kanduc2021shape}. We do not provide detailed answers to all the open questions but discuss possible starting points and avenues for further quantitative developments.  For this, we first take a look at polymer morphologies and describe how nanoclusters evolve in space. We then present challenges related to the low dielectric environment of polymers and the fact that those polymers that selectively transport ions are highly heterogeneous in terms of polymer and water domains.  Importantly, based on the solution--diffusion mechanism, our discussion involves two types of material parameters: (i) The solvation free energy of the ion species in the material, and (ii) the diffusion coefficients of the ion species~\cite{yaroshchuk2001non}. We clarify how solvation and partitioning can be characterized and interpreted by the free energy needed to transfer the ion from bulk solvent (e.g., water) into the material. We then turn to diffusion and explain activated hopping mechanisms and the role of water. We discuss how diffusion and solvation contribute to ion permeability and selectivity in dense polymer membranes. Finally, we bring our view in relation to some contemporary applications and challenges for membrane design.

\section{From nanodroplets to water channels}

Water molecules can penetrate various nanoporous materials, ranging from liquids, soft polymers, solid minerals (e.g., zeolites) to biological structures (e.g., plant vessels and wood)~\cite{kusoglu2017new, jiang2018low, santoro2019insertion, medeiros2019characterization, chen2019wood}.
Yet, a prerequisite for water to do so is a sufficient amount of hydrophilic groups in the material. 
Sorption (i.e., uptake) of water strongly modifies the material properties and 
is hence of great relevance for chemistry, materials science, and Earth science. However, in this article, we will focus solely on polymers.
A famous example of a neutral polymer in applications of hydrogels (hydrated polymer networks)
is PNIPAM, which will be central to our discussion. PNIPAM undergoes a sharp transition from good to poor solvent conditions upon heating, thus making it a thermoresponsive polymer~\cite{halperin2015poly}. It has a hydrophobic backbone and the polar amide group (--CO--NH--) in its sidechains, which gives the polymer its tunable hydrophilic character. 
On the other hand, charged groups are found in ionomers, such as Nafion, a sulfonated tetrafluoroethylene, developed by DuPont in the '70s~\cite{kusoglu2017new}. This ionomer also has a hydrophobic backbone, whereas the sidechains are terminated by charged sulfonate groups (--SO$_3^{-}$).





Water uptake in hydrophilic polymers is typically a multistep and multiscale process, driven by complex interactions between water, the hydrophilic, and the hydrophobic domains in the polymer.
How much water does a polymer take up depends on many parameters, such as water activity (tuned by humidity or dissolved solutes), temperature, and the chemical composition of the polymer. Very generally, the uptake is larger for higher water activity and for polymers with more hydrophilic groups and fewer cross-linkers~\cite{aryal2018impact, xu2020molecular}. 

\begin{figure}[h]\begin{center}
\begin{minipage}[b]{0.49\textwidth}\begin{center}
\includegraphics[width=\textwidth]{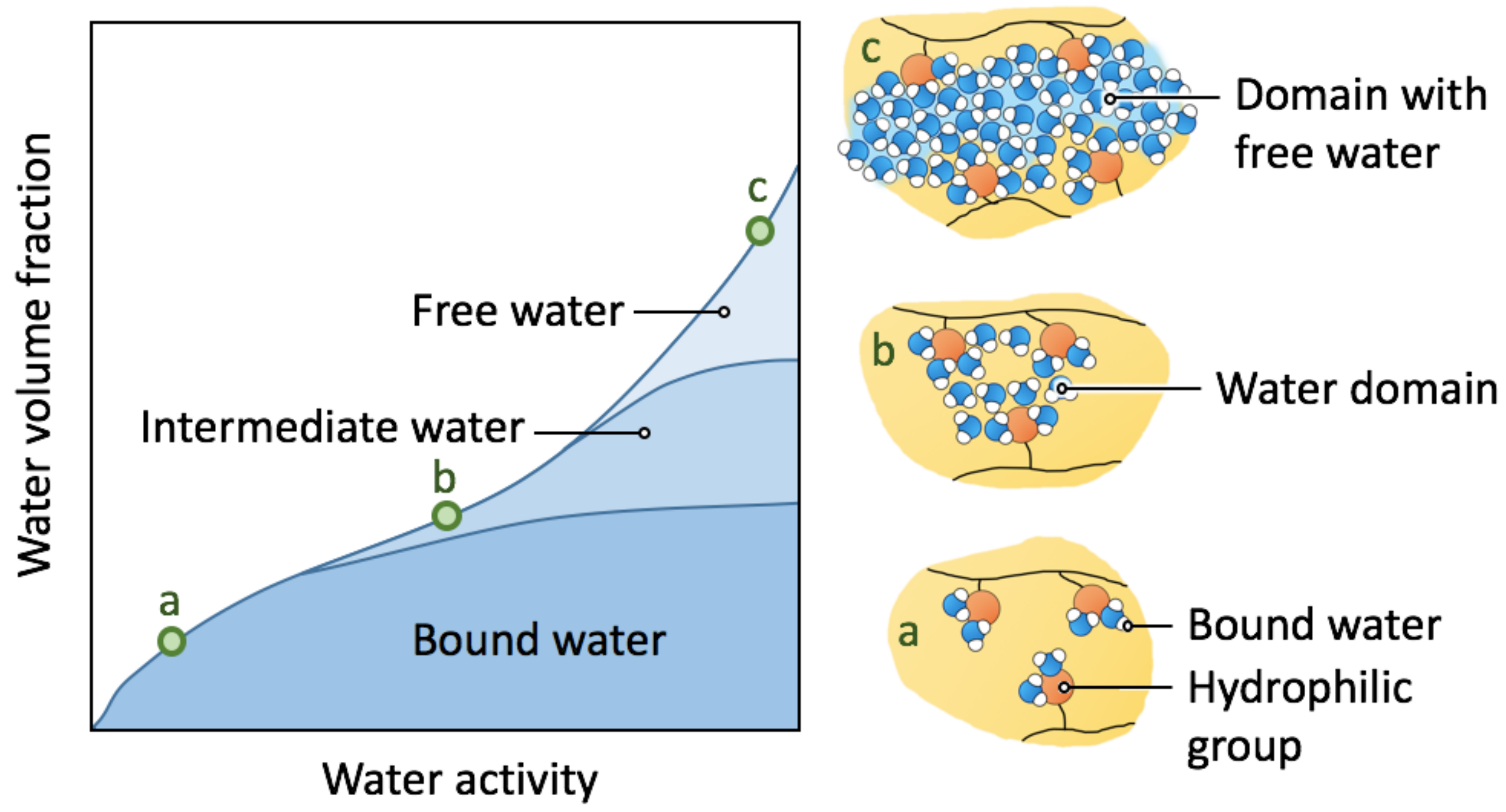}
\end{center}\end{minipage}
\caption{
Generic sorption isotherm of a hydrophilic polymer, indicating the amounts of (a) bound, (b) intermediate, and (c) free water. Depictions on the right show the corresponding growing water domains around hydrophilic groups.}
\label{fig:uptake}
\end{center}\end{figure}



\begin{figure*}[t]\begin{center}
\begin{minipage}[b]{0.96\textwidth}\begin{center}
\includegraphics[width=\textwidth]{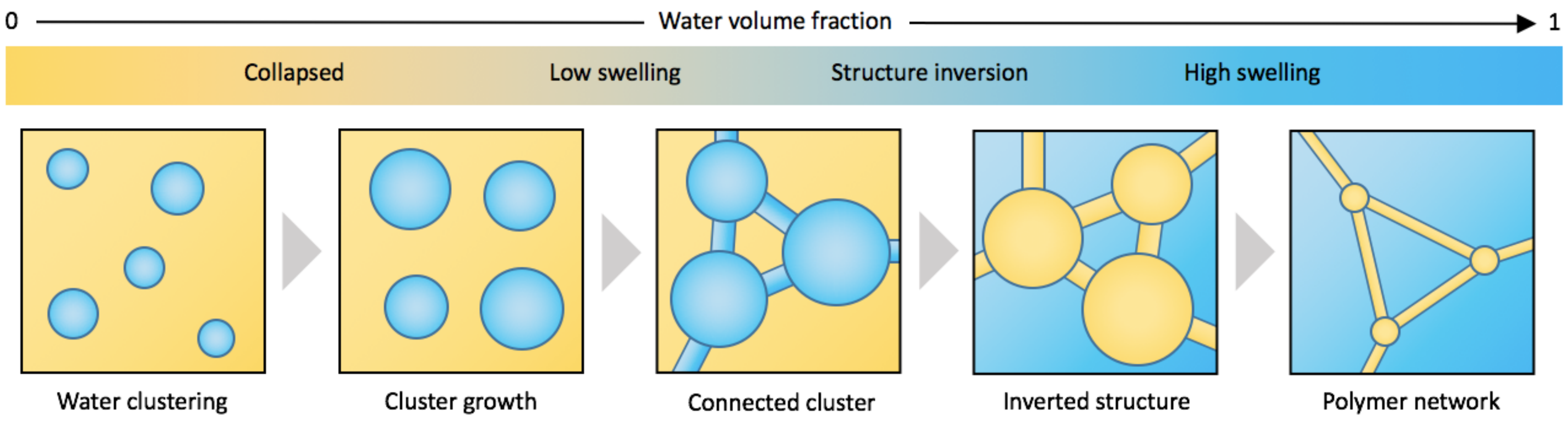}
\end{center}\end{minipage}
\caption{Structural idealization of polymer regimes upon hydration, showing dry polymer domains (yellow) and water domains (blue).
At very low hydration (low water packing fractions in the polymer), water forms individual nanosized domains in the form of droplets or clusters throughout the phase, which grow with increasing hydration. At even higher hydration, they connect and form a network, which goes on to an inverted structure and finally to a polymer network in water at very high hydration levels (high water volume fraction, close to unity). The figure was inspired by a similar one in Ref.~\cite{kusoglu2017new}.
}
\label{fig:regimes}
\end{center}\end{figure*}


A generic sorption isotherm---the amount of sorbed water versus the water activity---is shown in \Fig~\ref{fig:uptake}. 
In the `70s, Jhon and Andrade introduced a three-state classification of the water structure inside hydrogels based on their observations, which remains a useful concept up to this day~\cite{jhon1973water}: (a) ‘‘Bound’’ water is formed by water molecules that interact directly and strongly with primary hydrophilic sites, such that it behaves dynamically and thermodynamically as a part of the polymer chains. (b) ‘‘Intermediate’’ water consists of water molecules with weaker interaction with polymeric chains.  Finally, (c) ‘‘free’’ water is formed by water molecules with negligible interactions with polymer chains and retains the properties of bulk water.

A simplified and idealized view on different morphological regimes of water in polymers is sketched in \Fig~\ref{fig:regimes}.
At very low hydration levels, water molecules localize around hydrophilic groups and form small water domains---a kind of nanoclusters or nanodroplets. 
With increasing hydration, the clusters grow and start connecting with each other. 
The water morphology eventually undergoes a percolation transition from isolated water clusters to a three-dimensional interconnected network of water channels. 
Once the water amount becomes excessive, we can speak of an inverted structure, ultimately leading to a swollen polymer network in water, as the limiting scenario.

Our understanding of the transport of small molecules (i.e., much smaller than the mesh size of the network) in swollen networks is generally better than that in poorly hydrated, collapsed states. 
Transport in swollen states, featuring large amounts of water, can be more or less successfully described by various continuum, perturbative approaches~\cite{amsden1998solute}. In contrast, transport in poorly hydrated states is usually much more sensitive to the molecular architecture of the polymer and penetrants.
This means that already tiny chemical modifications in the structure can change the transport properties enormously. Yet, precisely this trait gives low hydrated materials their ability to be highly selective and favor passing certain kinds of penetrants over the others~\cite{lu2019tuning, epsztein2019activation, kanduc2021shape}.

\begin{figure*}[t]\begin{center}
\begin{minipage}[b]{0.62\textwidth}\begin{center}
\includegraphics[width=\textwidth]{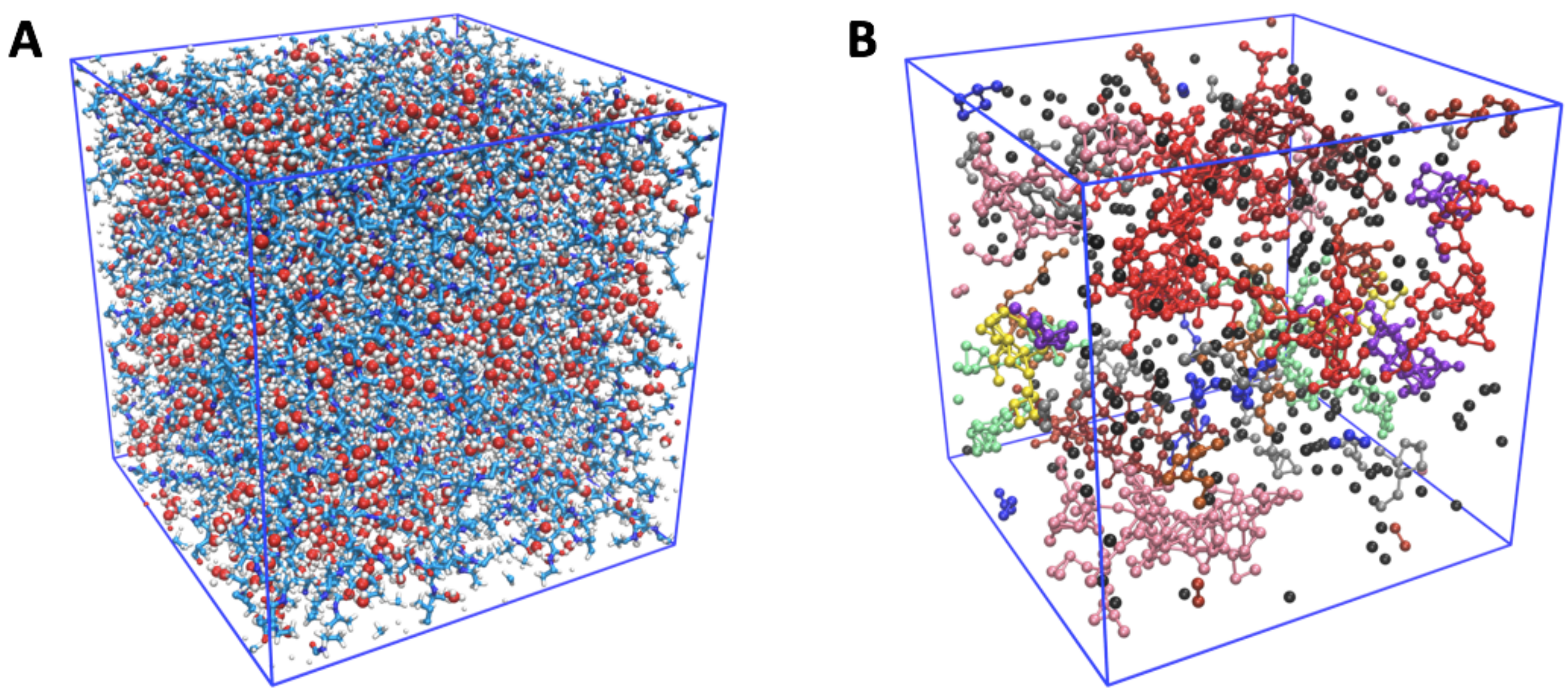}
\end{center}\end{minipage}\hspace{2ex}
\begin{minipage}[b]{0.28\textwidth}\begin{center}
\includegraphics[width=\textwidth]{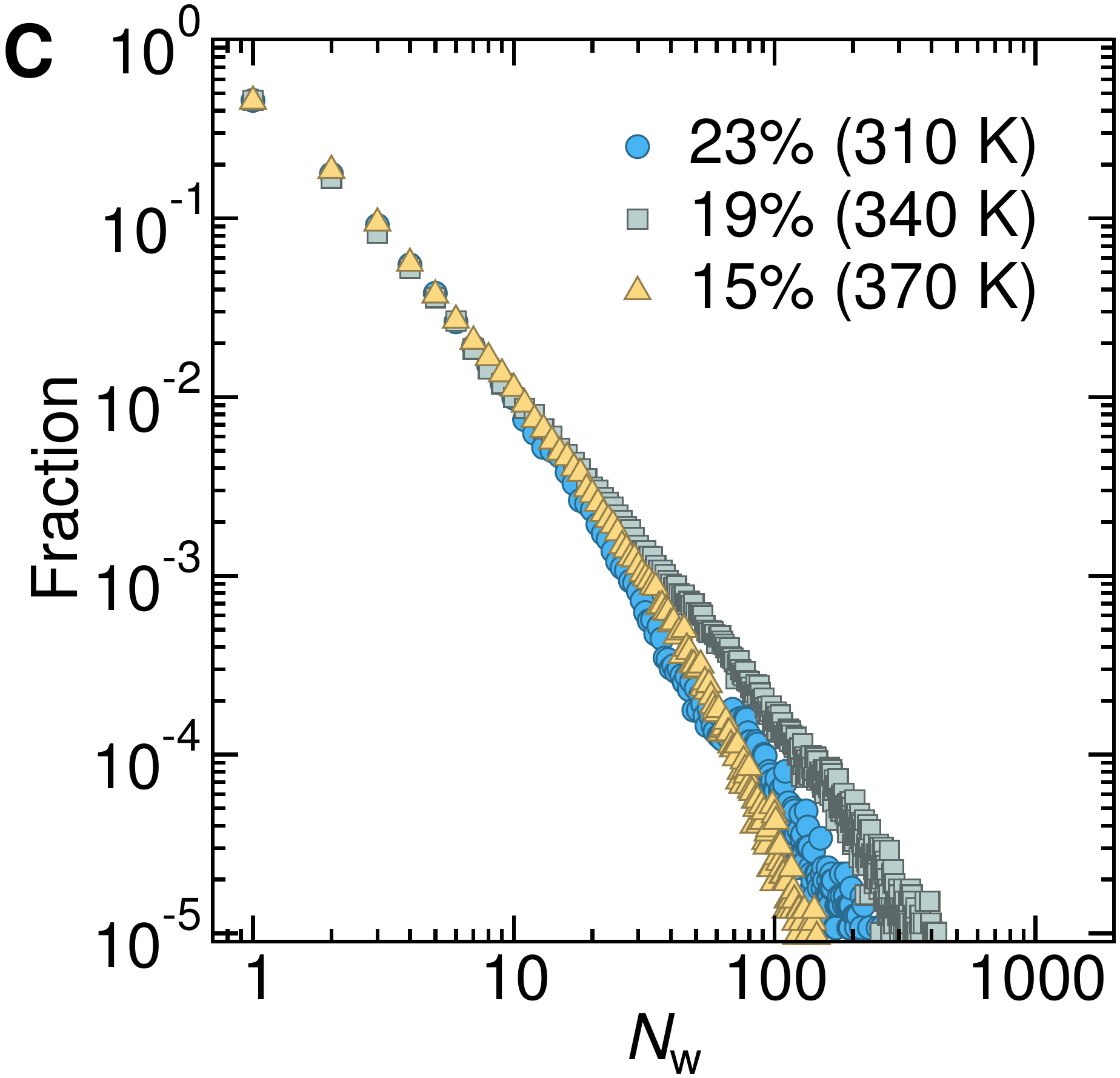}
\end{center}\end{minipage}
\caption{
(A) Molecular dynamics simulation snapshot of PNIPAM with 19 wt\% of water. (B) The same configuration showing individual water clusters distinguished by different colors (shown as connected water oxygen atoms). Panels (A)  and (B) reprinted with permission from Ref.~\cite{kanduc2018diffusion}, copyright 2018 American Chemical Society.
(C) Size distribution of water clusters in the collapsed polymer at three different water fractions and temperatures (water fractions result from chemical equilibrium with bulk water). Data points taken from Ref.~\cite{kanduc2018diffusion}.
}
\label{fig:clusters}
\end{center}\end{figure*}

In the rest of the paper, we focus exclusively on collapsed, low hydrated polymer states. 
Water domains in these states have been identified in numerous computer simulations, an example of which is shown in \Fig~\ref{fig:clusters}A for a dense PNIPAM polymer structure containing around 20 wt\% of water---thus mimicking a collapsed hydrogel at high temperature~\cite{kanduc2018diffusion}.

Individual nanosized water clusters (each one depicted in a different color in \Fig~\ref{fig:clusters}B) are far from being compact structures but rather of lacy, fractal-like forms. Their radius of gyration roughly follows a square-root dependence on the number of containing water molecules, $R_\trm{g}\propto N_\trm{w}^{1/2}$ for small clusters, which resembles a random walk~\cite{kanduc2018diffusion}. Besides, the clusters are polydisperse~\cite{kanduc2018diffusion, mabuchi2018relationship}, approximately following a power-law distribution, as shown in \Fig~\ref{fig:clusters}C.
The formation of individual clusters can be understood as a competition between water--water interactions (favoring two-phase separation) and the interactions between water and the hydrophilic polymer groups (favoring dispersion of water molecules).  
In an entirely nonpolar material (such as oil), the water completely phase separates from the rest of the material, ending up in the form of one single water drop. 

\section{Partitioning of ions in heterogeneous membranes}
We now turn our attention to the question of how easily ions can enter a polymer material that features low hydration and thus a low dielectric environment.
It has long been known that nonpolar or weakly polar media, such as poorly hydrated polymers or lipid bilayers, act as a barrier to the passage of ions between two aqueous solutions.
Since the electrostatic interaction is long-ranged, the leading term in the solvation free energy of a charged species in a given medium can be estimated within a continuum dielectric description~\cite{parsegian1969energy}.



For a spherical elementary charge $e$ of radius $a$
in an infinitely large medium of dielectric constant $\varepsilon_i$, the electrostatic self-energy is expressed in terms of the Born charging energy as $G_\trm{B}=e^2/(8\pi\varepsilon_i\varepsilon_0 a)$, where $\varepsilon_0$ is the vacuum permittivity.
Thus, the work needed to transfer the charge from water (of dielectric constant $\varepsilon_\trm{w}=80$) into a polymer phase (of dielectric constant $\varepsilon_\trm{p}$) is equal to
\begin{equation}
\Delta G_\trm{B}=\frac{e^2}{8\pi\varepsilon_0 a}\left(\frac{1}{\varepsilon_\trm{p}}-\frac{1}{\varepsilon_\trm{w}}\right)
\label{eq:dGsphere}
\end{equation}
and is referred to as the Born transfer free energy. To get the feeling about the energy scale, let us consider a monovalent ion of radius $a=0.25$~nm and a pure hydrocarbon material (e.g., oil) with $\varepsilon_\trm{p}=2$. Equation~\ref{eq:dGsphere} then amounts to a considerable value of $\Delta G_\trm{B}=55\,\kB T$, where $\kB T$ is the thermal energy---$\kB$ being the Boltzmann constant and $T$ the temperature.
In this simple Born solvation picture, one can immediately realize that a low dielectric constant, as encountered in hydrocarbon materials, leads to notable electrostatic penalties for ions.
\begin{figure*}[t]\begin{center}
\begin{minipage}[b]{0.66\textwidth}\begin{center}
\includegraphics[width=\textwidth]{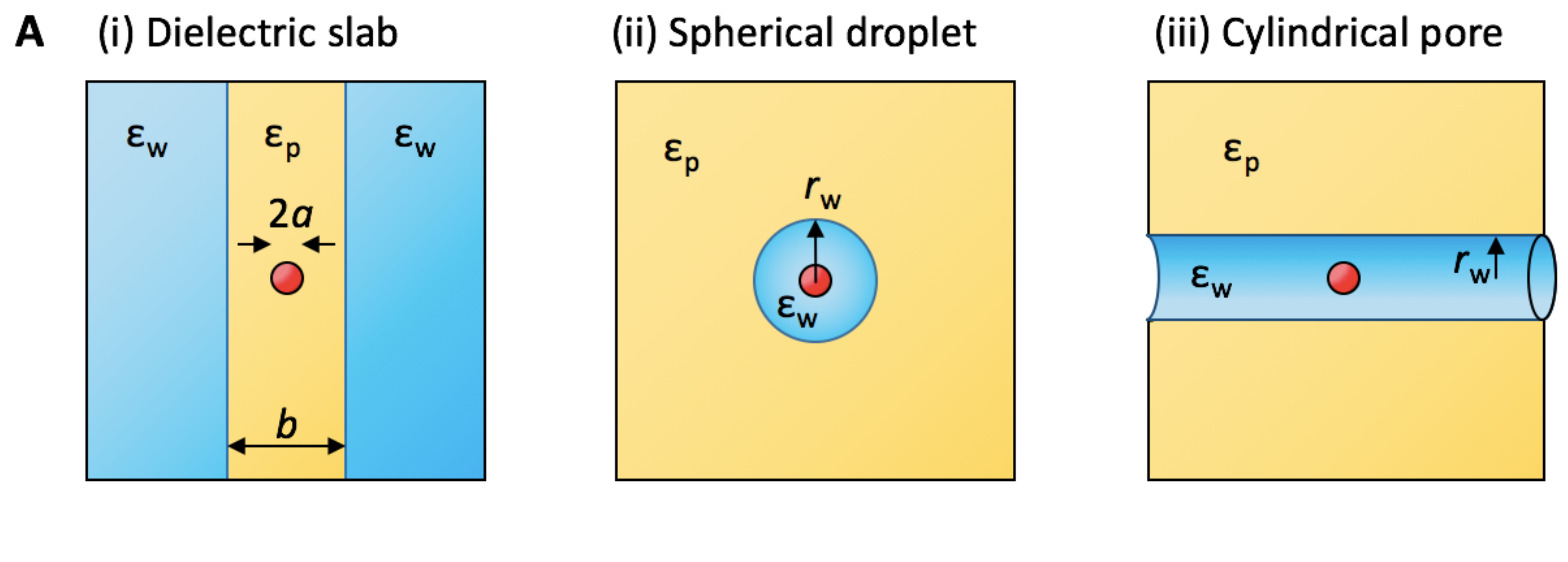}
\end{center}\end{minipage}\hspace{3ex}
\begin{minipage}[b]{0.3\textwidth}\begin{center}
\includegraphics[width=\textwidth]{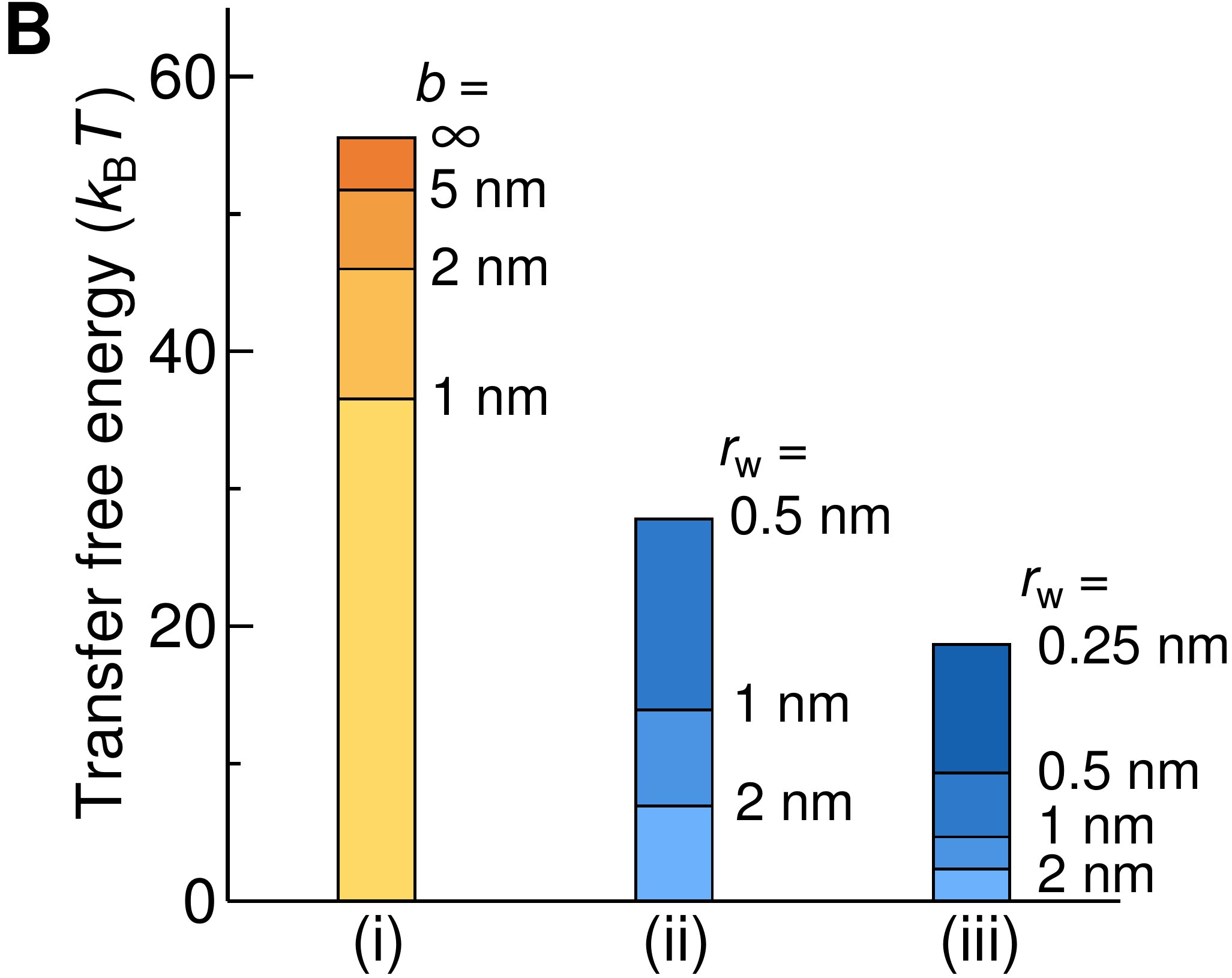}
\end{center}\end{minipage}
\caption{(A) Various ways by which an ion of radius $a$ can enter a dielectric medium (yellow) of dielectric constant $\epsilon_p$. The blue regions depict water with dielectric constant $\epsilon_w$; see text for details. (B) Calculated Born transfer free energies from bulk water into the configuration shown in (A), assuming $\varepsilon_\trm{w}=80$, $\varepsilon_\trm{p}=2$, and $a=0.25$~nm~\cite{parsegian1969energy}.
}
\label{fig:energies}
\end{center}\end{figure*}
There are, however, important factors that lower the above estimate in various cases 
dealing with ions crossing low dielectric materials.
Prototypical scenarios of ion crossings were analyzed by Parsegian more than half a century ago~\cite{parsegian1969energy}, which we briefly recap in the following, see \Fig~\ref{fig:energies}.

In the first scenario, the low dielectric medium is a planar slab of thickness $b$, bounded on two sides by semi-infinite regions of water (\Fig~\ref{fig:energies}A(i)).
The finite thickness of the slab material causes the free energy penalty to decrease because of the high dielectric constant of water $\varepsilon_\trm{w}$ outside. For an ion at the center of the slab (where the penalty is the highest), the decrement to the infinite-slab transfer free energy (i.e., \Eq~\ref{eq:dGsphere}) due to the finite thicknesses is
\begin{equation}
\Delta\Delta G=-\frac{e^2}{4\pi\varepsilon_\trm{p}\varepsilon_0 b}\,\ln\frac{2\varepsilon_\trm{w}}{\varepsilon_\trm{w}+\varepsilon_\trm{p}}
\label{eq:GB}
\end{equation}
Figure \ref{fig:energies}B shows the calculated transfer free energy for several different slab thicknesses and reveals that the influence of the finite thickness is negligible for membranes more than several nanometers across.


However, an ion may not get entirely rid of its hydration shell but remains entrapped in a water shell, which acts as a ``carrier''. In the simplest view, the hydration shell can be represented as a spherical water droplet (\Fig~\ref{fig:energies}A(ii)) of radius $r_\trm{w}$, such that the transfer free energy from the water phase into the center of the droplet is
\begin{equation}
\Delta G_\trm{B}=\frac{e^2}{8\pi\varepsilon_0 r_\trm{w}}\left(\frac{1}{\varepsilon_\trm{p}}-\frac{1}{\varepsilon_\trm{w}}\right)
\label{eq:GBdroplet}
\end{equation}
This expression is essentially the same as \Eq~\ref{eq:dGsphere}, but with the bare ion radius $a$ replaced by the droplet radius $r_\trm{w}$, and precisely this detail substantially reduces the free energy. As depicted in \Fig~\ref{fig:energies}B, the hydration shell of radius 0.5~nm reduces the free energy 2-fold compared with the completely dehydrated scenario, whereas a shell of a one-nm-radius yields a 4-fold reduction.

Finally, an ion can pass a low-dielectric medium through a water pore or channel.
The channel can be represented as a very long cylinder of radius $r_\trm{w}$ and the dielectric constant of water $\varepsilon_\trm{w}$ (\Fig~\ref{fig:energies}A(iii)). With the surrounding dielectric constant of $\varepsilon_\trm{p}$, the work for transferring a charge from bulk water into the middle of the cylinder is
\begin{equation}
\Delta G_\trm{B}=\frac{e^2}{4\pi\varepsilon_p\varepsilon_0 r_\trm{w}} F\left(\frac{\varepsilon_\trm{p}}{\varepsilon_\trm{w}}\right)
\end{equation}
The dimensionless function $F$ should be calculated numerically (see, e.g., Refs.~\cite{parsegian1969energy, cui2006electrostatic}). For $\varepsilon_\trm{p}/\varepsilon_\trm{w}=2/80$, its value is around $F\approx 0.165$~\cite{parsegian1969energy}.
For the cylindrical channel, similarly as for the droplet carrier, the transfer free energy is inversely proportional to the pore radius and is consequently much lower than that for the bare ion (see \Fig~\ref{fig:energies}B).

Another relevant effect that reduces the electrostatic penalty is the increase of the global dielectric constant of the polymer due to water clusters.  While pure hydrocarbons typically feature $\varepsilon_\trm {p}\approx2$, introducing water into the polymer matrix gradually increases a dielectric constant. Clearly, in the limit of a highly swollen network, $\varepsilon_\trm{p}$ approaches the one of bulk water, $\varepsilon_\trm{w}$.
Collapsed PNIPAM hydrogels have the dielectric constant much below that of bulk water, but with a typical value of $\varepsilon_\trm{p}\approx10$, the transfer free energies in \Fig~\ref{fig:energies} get reduced by around a factor of 5.



\begin{figure*}[t]\begin{center}
\begin{minipage}[b]{0.36\textwidth}\begin{center}
\includegraphics[width=\textwidth]{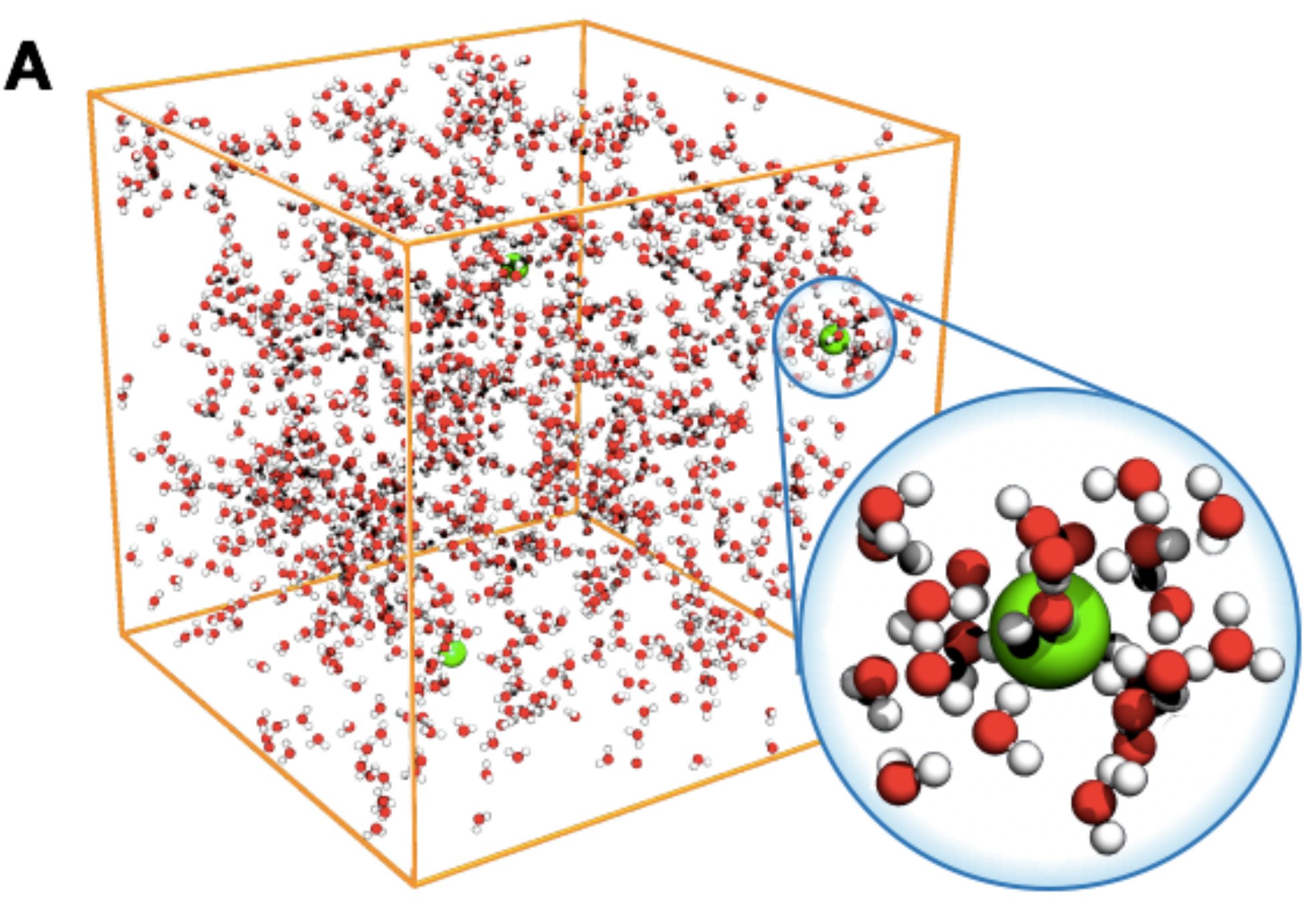}
\end{center}\end{minipage}\hspace{1ex}
\begin{minipage}[b]{0.25\textwidth}\begin{center}
\includegraphics[width=\textwidth]{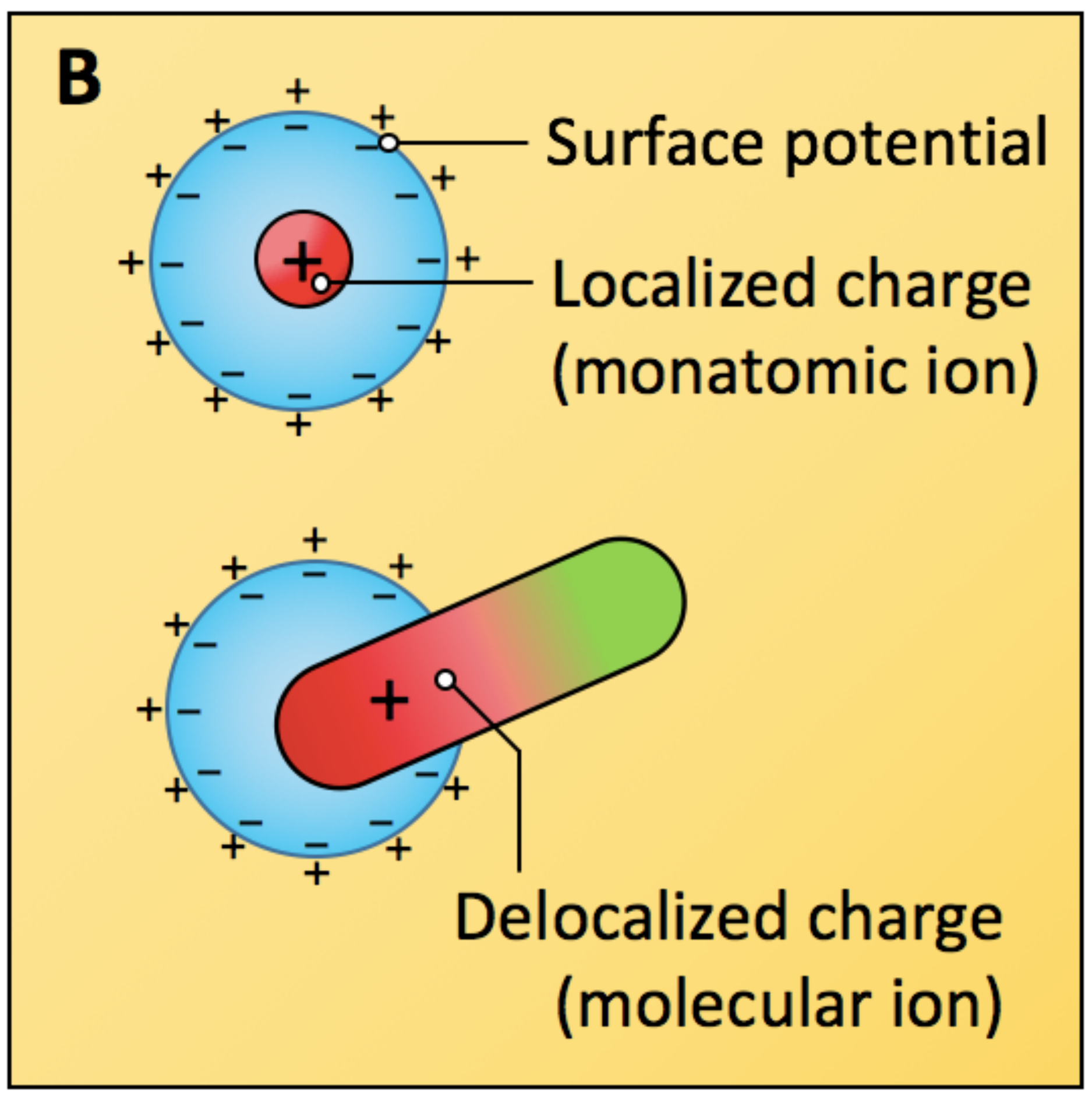}
\end{center}\end{minipage}\hspace{1ex}
\begin{minipage}[b]{0.35\textwidth}\begin{center}
\includegraphics[width=\textwidth]{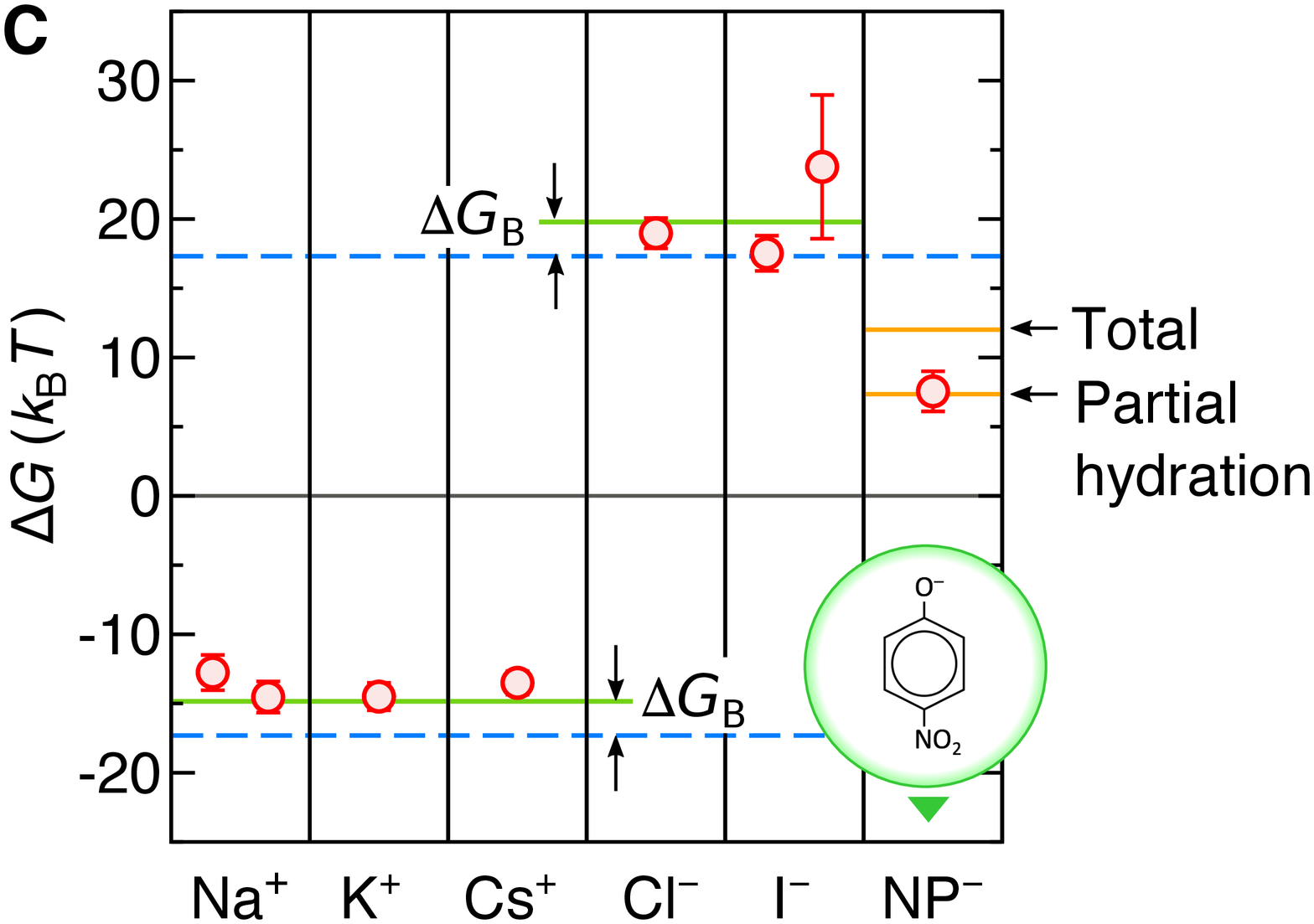}
\end{center}\end{minipage}
\caption{
(A) Snapshot of chloride ions in a PNIPAM phase (shown only water molecules and ions, no polymer). Reprinted with permission from Ref.~\cite{kanduc2019aqueous}, copyright 2019 American Chemical Society. 
(B)~Continuum picture of an ion encapsulated by a water cluster (top) and a larger molecular ion that partially sticks out of the hydrating cluster (bottom).
(C)~Transfer free energies of ions from water into collapsed PNIPAM obtained from simulations (red circles)~\cite{kanduc2019aqueous}. The blue dashed lines depict the contributions from the water interface potential $\pm e\psi_\trm{s}$, the green solid lines are the added Born prediction (\Eq~\ref{eq:GBdroplet}) of $\Delta G_\trm{B}=2.6\,\kB T$ for monatomic ions hydrated by water droplets of radius $r_\trm{w}=1$~nm. Orange lines are Born predictions~\cite{kanduc2019aqueous} for nitrophenolate in the case of full and the actual partial hydration (see text). Adapted with permission from Ref.~\cite{kanduc2019aqueous}, copyright 2019 American Chemical Society. 
}
\label{fig:ions}
\end{center}\end{figure*}

 Even though the above simple calculations are based on a continuum dielectric model and neglect molecular details, they are very illustrative and allow drawing this fundamental conclusion: 
 Removing the entire hydration shell from an ion is energetically too costly. Reasonable energies associated with the transfer of ions into low dielectric materials are inseparably linked to a hydration carrier~\cite{parsegian1969energy}.
 This notion has been confirmed and studied by many computer simulations in various contexts~\cite{kanduc2019free,duvail2019uo, widstrom2021water}. Figure~\ref{fig:ions}A shows a snapshot of chloride ions, hydrated with water, in a collapsed PNIPAM phase~\cite{kanduc2019free}.
Despite the irregular shapes of the water clusters, it nevertheless turns out that
a spherical approximation of clusters (as depicted in \Fig~\ref{fig:ions}B top) is
good enough for simple estimates involving monatomic ions~\cite{kanduc2019aqueous}. 
Things get more complicated, however, with some molecular ions that are not well hydrated, as we will discuss later on.


A subtle ingredient to the story of ionic solvation, not considered in the Born solvation picture, is the water interface potential.  
The water interface at a nonpolar medium (e.g., air or hydrocarbon) acquires an electrostatic potential stemming from the ordering of water dipoles at the boundary. 
There is a lack of consensus about the value of the interface potential, yet classical simulations typically give the value of around $\psi_\trm{s} \approx -$0.5 V with respect to the surrounding nonpolar medium~\cite{vacha2011orientation, caleman2011atomistic, beck2013influence, kanduc2019free}.
It turns out that this potential is unimportant in the vast majority of cases. When an ion enters a nanocluster from a bulk water phase, it crosses two water boundaries, and with that, the two opposing contributions from the surface potential cancel one another. However, if the potential at the macroscopic water--polymer interface is screened by ions, only the potential jump at the nanocluster remains. The interface potential contribution from water nanoclusters can be observed in simulations for single-ion transfer free energies, as shown in \Fig~\ref{fig:ions}C. There, the distinction between cations and anions (red circles) is primarily due to the water potential of nanoclusters, $\pm e\psi_\trm{s}=\pm 17\,\kB T$ (indicated by blue dashed lines)~\cite{kanduc2019aqueous}. The estimated Born free energy from \Eq~\ref{eq:GBdroplet} for a spherical droplet with $r_\trm{w}=1$~nm and $\varepsilon_\trm{p}=8.5$ (estimated from the simulations~\cite{kanduc2019aqueous}) is 
  $\Delta G_\trm{B}=2.6~\kB T$, and is added on top of the interface potential contributions as green solid lines in \Fig~\ref{fig:ions}C.
Despite the significant contribution of the water interface potential to the single-ion transfer free energies, it has, on the other hand, no influence on the final concentrations of fully hydrated ions in the thermodynamic limit---namely, the same number of cations and anions are enclosed by water clusters and the opposing contributions from the interface potential cancel out.  

The above notion of ion solvation also applies to the uptake of salt. 
The uptake is typically quantified by the partition ratio, $K_\trm{salt}$, defined as the concentration ratio of ions inside and outside the polymer.
For the simplest case of 1:1 electrolyte (such as sodium chloride), the partition ratio is related to the transfer free energies as
\begin{equation}
K_\trm{salt}=\rme^{-\beta\left(\Delta G^{(+)}+\Delta G^{(-)}\right)/2}
\label{eq:Ksalt}
\end{equation}
where $\Delta G^{(+)}$ and $\Delta G^{(-)}$ are the transfer free energies of  cations and anions, respectively, from water into the polymer material. As seen, salt partitioning is a collective effect, dependent on the sum of transfer free energies of the two ion species, and clearly shows how the interface potential contributions $e\psi_\trm{s}$ and $-e\psi_\trm{s}$, depicted in \Fig~\ref{fig:ions}C, cancel out.

\begin{figure}\begin{center}
\begin{minipage}[b]{0.37\textwidth}\begin{center}
\includegraphics[width=\textwidth]{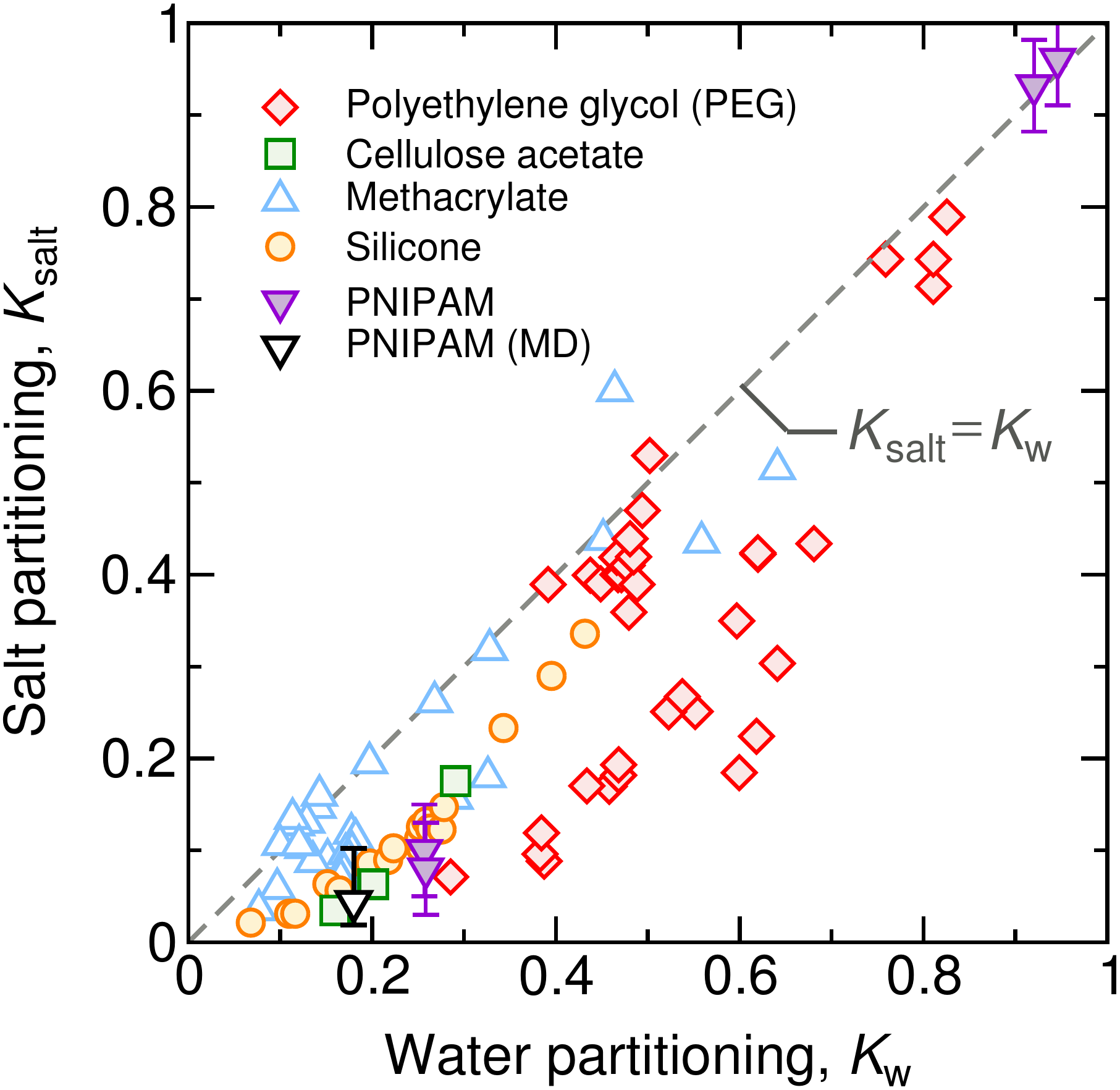}
\end{center}\end{minipage}
\caption{Partition ratio of NaCl versus the water partition ratio for several polymers
at different temperatures or with different degrees of copolymerization as measured experimentally (for references, see Ref.~\cite{kanduc2019aqueous}) and obtained from MD simulations of PNIPAM~\cite{kanduc2019aqueous}).
Adapted with permission from Ref.~\cite{kanduc2019aqueous}, copyright 2019 American Chemical Society. 
}
\label{fig:Ksalt}
\end{center}\end{figure}

Figure \ref{fig:Ksalt} shows the correlation between the sodium chloride salt partitioning ($K_\trm{salt}$) and water partitioning ($K_\trm{w}$; defined as the ratio of water densities inside and outside the polymer) for a few uncharged hydrogels (taken from literature, see references in Ref.~\cite{kanduc2019aqueous}).
The uptake of ions evidently depends on water amount: 
Polymers containing more water in general also sorb more salt than those polymers with less water.
The diagonal dashed line depicts an apparent limiting scenario of $K_\trm{salt} = K_\trm{w}$ for which the salt concentration in the sorbed water is equal to that in bulk water.
However, most data points are below the dashed line, implying that both polymer--ion and polymer--water interactions influence ion partitioning~\cite{kanduc2019aqueous}.
The single-ion transfer free energies for Na$^+$ and Cl$^-$ from our simulations in  \Fig~\ref{fig:ions}C result in the partitioning (using \Eq~\ref{eq:Ksalt}) indicated by the white triangle symbol in \Fig~\ref{fig:Ksalt}, which is in the ballpark of the experiments. Moreover, the Born model (with the estimated $\Delta G_\trm{B}=2.6\,\kB T$ for both ions; see \Fig~\ref{fig:ions}C) predicts $K_\trm{salt}=\exp(-\beta\Delta G_\trm{B})\approx 0.05$ (regardless of whether $\psi_\trm{s}$ is included or not), which is in good agreement with the simulation result. 
 


However, things get more involved when the symmetry between positive and negative charges of the hydrated parts of the molecule is broken. This occurs, for instance, when the electron charge of at least one ion species is delocalized (i.e., the charge is smeared over several atoms), such as in ionized conjugated molecules (e.g., aromatic compounds).
In this way, the charge density is lower and attracts water less strongly, which in turn can lead to partial dehydration of the charge. In such a case, not the entire molecule's charge is enclosed by a water cluster and subjected to the water interface potential, as schematically depicted in \Fig~\ref{fig:ions}B (bottom). The net contribution from the water interface potential is therefore non-zero, and it impacts the partitioning of ions. 

In our recent simulation study~\cite{kanduc2019aqueous}, we found that the molecular ion nitrophenolate (see its depiction in the circular inset of \Fig~\ref{fig:ions}C) is partially dehydrated. Consequently, its transfer free energy decreases---compare the estimated values for a hypothetically fully hydrated and the actually partially hydrated ion, indicated by the orange lines in \Fig~\ref{fig:ions}C.
This reduction of the transfer free energy by several $\kB T$ increases the partitioning by orders of magnitude. These results suggest that ionizing a molecule can, in fact, even boost the partitioning in collapsed, poorly hydrated hydrophobic gels in some cases rather than hinder it, which challenges the traditional simplistic view on ion solvation.
Moreover, molecular ions gain a significant contribution to the free energy from their neutral parts. This contribution scales very well with the solvent-accessible surface area~\cite{kanduc2019free, kanduc2021shape}.
The nonpolar parts of the molecules preferentially sorb in dry regions of the polymer, whereas polar parts are immersed inside the water nanodroplets~\cite{kanduc2019free}. A more thorough discussion on the hydration of molecular ions can be found in Ref.~\cite{kanduc2019aqueous}.

Entrapped ions in hydrophilic nanodomains can also be found in numerous other situations; one nice example is ion extraction from aqueous phases (such as ore processing and recycling). 
These procedures typically use amphiphilic extractant molecules that form self-assembled aggregates with hydrophilic nanodomains in the middle~\cite{duvail2019uo}. These nanodomains, which resemble water nanoclusters in polymers, can trap and hydrate ions from the aqueous solution and enable their removal from the aqueous phase~\cite{spadina2019synergistic,spadina2020multi}.

\section{Diffusion in poorly hydrated polymers}
\begin{figure*}[t]\begin{center}
\begin{minipage}[b]{0.5\textwidth}\begin{center}
\includegraphics[width=\textwidth]{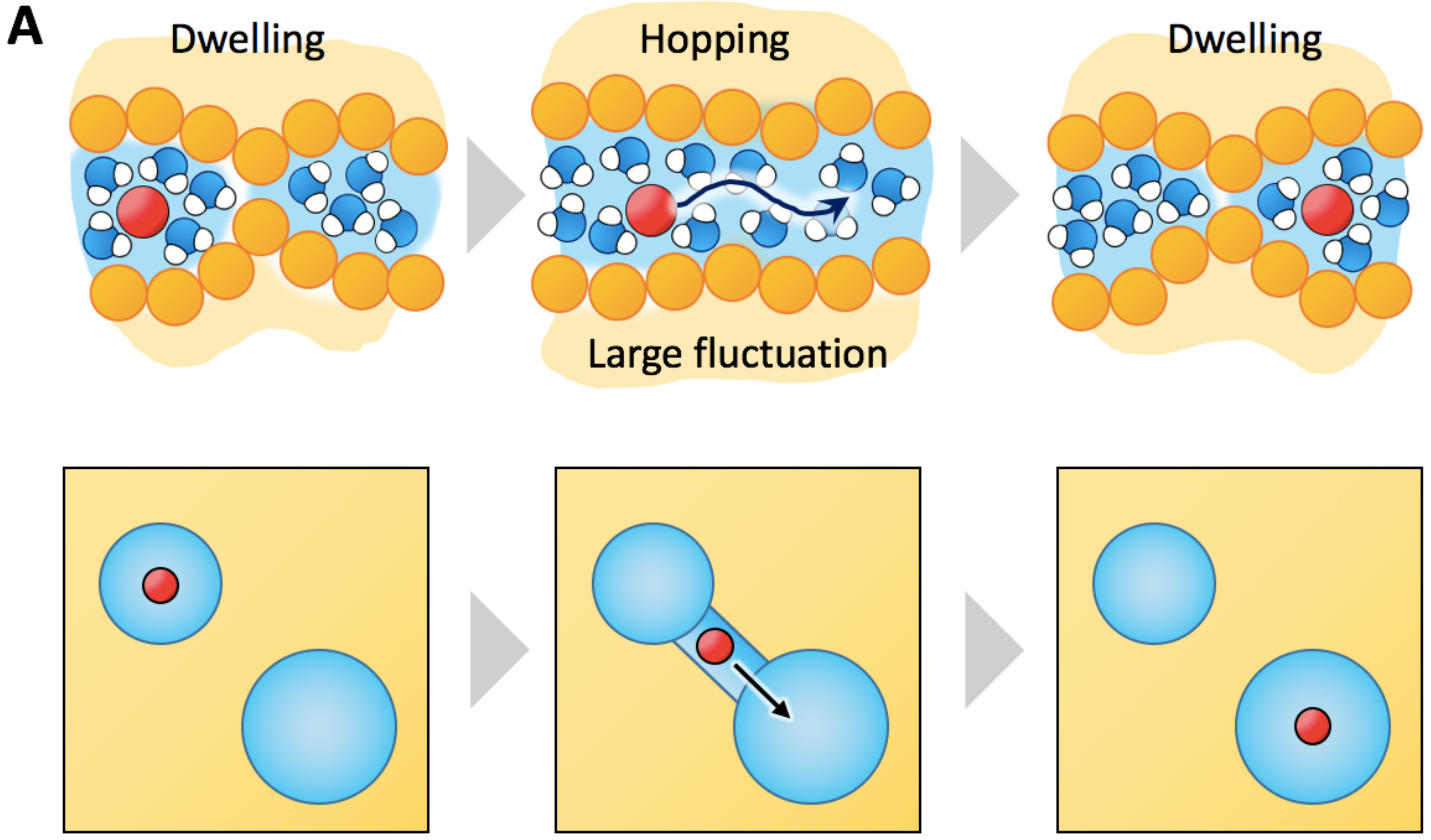}
\end{center}\end{minipage}\hspace{3ex}
\begin{minipage}[b]{0.33\textwidth}\begin{center}
\includegraphics[width=\textwidth]{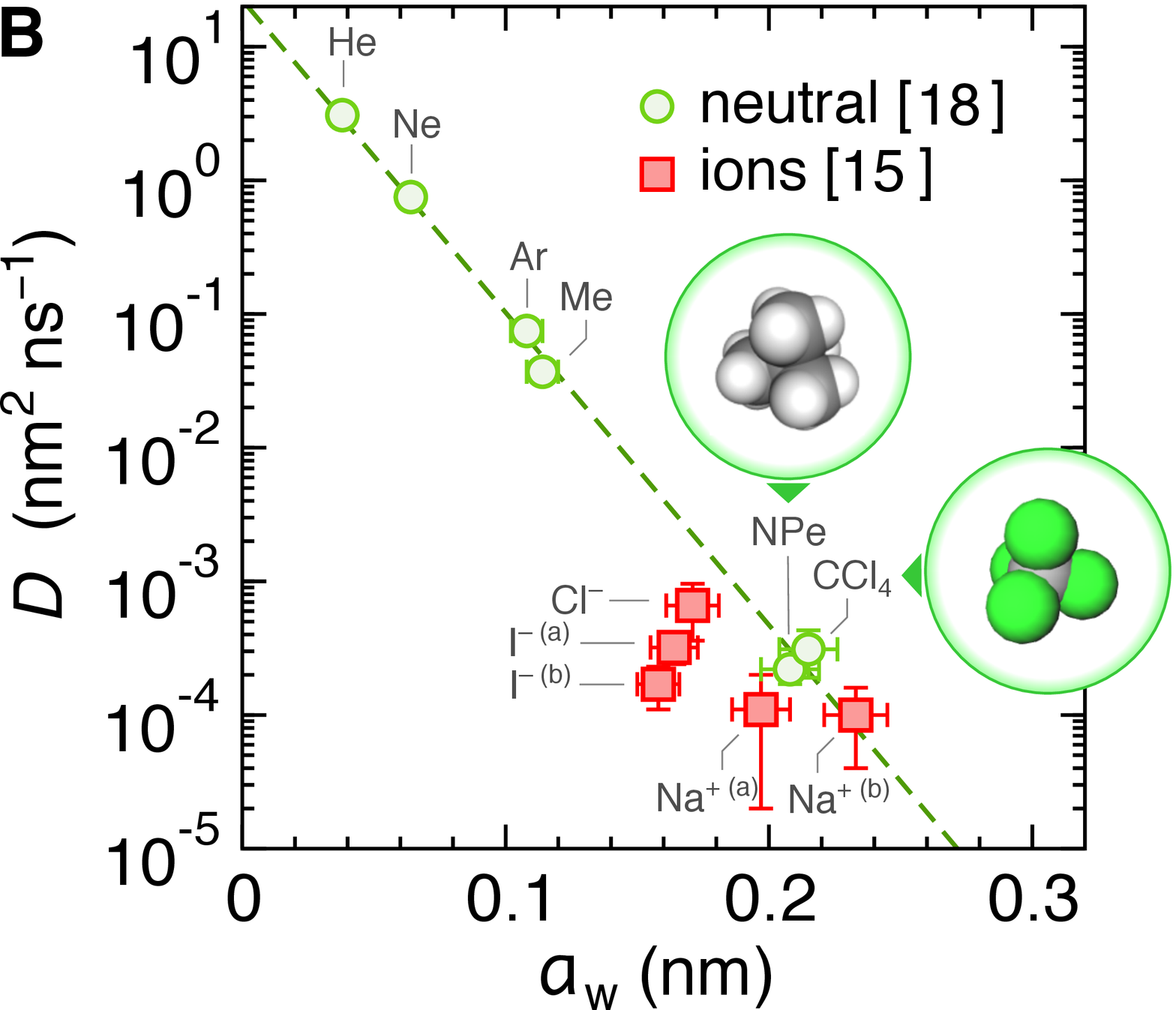}
\end{center}\end{minipage}
\caption{(A) Schematic depiction of hopping diffusion in the presence of water on the molecular level (top) and the continuum level (bottom).
(B) MD simulation results for diffusion coefficients of spherical neutral penetrants [circles: helium (He), neon (Ne), argon (Ar), methane (Me), neopentane (NPe), and tetrachloromethane (CCl$_4$)]~\cite{kanduc2021shape} and monovalent monatomic ions (squares; \cite{kanduc2018diffusion}) 
  in a collapsed PNIPAM polymer matrix versus their Stokes radii in bulk water.
  For Na$^+$ and I$^-$ data from two different force fields were used; see Ref.~\cite{kanduc2018diffusion} for more details. Adapted with permission from Ref.~\cite{kanduc2021shape}, copyright 2021 American Chemical Society.  
}
\label{fig:diffusion}
\end{center}\end{figure*}

The other necessary quantity for understanding ionic transport is the diffusion coefficient.
Diffusion in dense polymer systems is a highly complex and frequently debated topic. 
It differs significantly from Brownian diffusion in simple liquids and is featured by various mechanisms and regimes, depending on material and environmental parameters, such as polymer volume fraction, penetrant size, and temperature~\cite{zhang2018coarse}.
In dense polymers, penetrants most of the time dwell in a local cavity, trapped by surrounding polymer chains.
A large enough thermal fluctuation creates a short-lived channel between the polymer chains into which the highly confined penetrant can jump and propagate to a new location, where it then dwells again for some time, as depicted in \Fig~\ref{fig:diffusion}A.
This, the so-called, hopping mechanism has been revealed in computer simulations of polymer melts (i.e., without water) in the `90s~\cite{takeuchi1990jump, muller1991diffusion}.

However, it is less known what role water plays in hopping diffusion in collapsed hydrated polymers, such as PNIPAM, for instance. Once a channel is created, water in the polymer can ``flood'' the created passage. It is known that water can wet pores and cavities of nanoscopic dimensions, such as those in nanotubes, proteins, and ion channels, sometimes even in as a single-file hydrogen-bonded wire~\cite{rasaiah2008water, brewer2001formation, dellago2003proton}. 
In addition to that, nonpolar pores form excellent low-friction conduits for the flow of water~\cite{rasaiah2008water}.
The reason is that as water molecules pass through the pore, they do not form strong interactions with the pore, and therefore do not transfer translational momentum to it.

These characteristics play a key role in the diffusion of penetrant molecules in dense hydrogels.
It is, therefore, no surprise that small penetrants, regardless of being polar or nonpolar, charged or neutral, travel through water channels rather than diffusing through dry parts of the polymer~\cite{kanduc2018diffusion}. 
This is because the transient channels, which are the primary pathway for transportation by hopping, are inevitably filled with water.

Hopping diffusion is an activated process since the creation of a pore relies on a large enough thermal fluctuation. Consequently, the diffusion coefficient scales as $D=D_0 \exp(-\Delta F_\trm{a}/\kB T)$, where $\Delta F_\trm{a}$ is the free energy for creating the pore. 
Nonetheless, a so-far unresolved conundrum is, how does $\Delta F_\trm{a}$ scale with the pore radius, or equivalently, the radius of the penetrant, $a_\trm{w}$, that passes through?
Most of the established theories and computer simulations of polymer melts (i.e., without solvent) in the rubbery regime as well as (implicit solvent) coarse-grained simulations suggest a square dependence, $\Delta F_\trm{a}\propto a_\trm{w}^2$, or even a cubic dependence $\Delta F_\trm{a}\propto a_\trm{w}^3$.
However, the understanding has been challenged by our recent studies of a hydrated PNIPAM polymer, which convincingly demonstrated a linear scaling $\Delta F_\trm{a}\propto a_\trm{w}$~\cite{kanduc2018diffusion, kanduc2021shape}.
In other words, diffusion in a system containing water is faster for larger penetrants than in dry systems. It is, nevertheless, widely known that water in polymers acts as a plasticizer and softens the polymer matrix, which also eases the diffusion of small molecules. 
A theoretical explanation in this direction is offered by recent theoretical concepts by Schweizer and coworkers, who showed that a coupled dynamics in dense liquids indeed results in a linear size dependence of the free energy barrier~\cite{zhang2017correlated, mei2021activated}.

Our knowledge of diffusion becomes even more obscured when it comes to ions.
Figure \ref{fig:diffusion}B compares diffusion coefficients of neutral spherical molecules and monovalent monatomic ions in the collapsed PNIPAM model~\cite{kanduc2018diffusion, kanduc2021shape}. Spherical neutral penetrants (ranging by size from helium, neon, methane, neopentane, to tetrachloromethane), plotted by green circles, clearly follow the relationship $D=D_0 \exp(-a_\trm{w}/\lambda)$, depicted by a dashed line. Nonetheless, ions (plotted by red squares) significantly deviate from the trend of neutral penetrants and do not seem to exhibit a well-defined trend. 
All we can conclude from the plot is that ions diffuse more slowly than neutral molecules of a similar size.

The mechanisms for ion diffusion in PNIPAM polymer membranes have so far not been scrutinized, which prevents us from offering a firm explanation. Based on the current more general understanding of ion diffusion through membranes~\cite{epsztein2020towards}, we can only speculate on several reasons for the slower diffusion.
The first one is that ions maintain a sizable hydration shell. If a transient channel is too narrow, the ion cannot pass readily through because it would need to shed a significant fraction of its hydration shell~\cite{epsztein2019activation}, as also we already concluded from the continuum dielectric approach,  \Fig~\ref{fig:energies}B.
The second reason is that ions interact with polar groups of the polymer additionally with strong Coulomb forces, thus making the energy landscape rougher, which in turn slows down diffusion~\cite{kim2019tuning}. In contrast, Coulomb forces are absent in neutral penetrants, whose diffusion is primarily governed by the steric effects between the penetrant and the polymer matrix~\cite{zhang2017correlated, kanduc2021shape}.
Another possible {\it modus operandi} is that ions are only transported by the slow diffusion of the water clusters themselves. Probably the various mechanisms are all operational and balanced by system-specific features, foremost the hydration level.

Figure \ref{fig:diffusion}B also implies a high ion specificity in membrane diffusion, that is, tiny details in the ionic interactions with water and polymers impact the diffusion, unlike for neutral solutes. This is a noteworthy observation for filtration and selective permeability in dense polymers and requires further investigation.



\section{Transport through membranes}

From the two material parameters discussed in the previous two sections, partitioning and diffusivity, it is possible to quantify the transport of ions based on established electrodiffusion theories, such as the Poisson--Nernst--Planck equation, for instance~\cite{yaroshchuk2001non, graf2000dynamic, lu2010poisson}.
The transport can be quantified in various ways, such as by conductance or, even more generally, by permeability. Permeability is a measure of how easily an ion (or any other molecule) can cross a material.
A flux of ions, driven either by a concentration gradient or electric field, is proportional to the permeability $P$~\cite{yaroshchuk2001non}.
In general, permeability can be quite a complex function of diffusivities and partitionings of all ionic species involved in the electrolyte.

Nevertheless, in simple cases in which a charged penetrant species is highly dilute and electrostatically screened by background salt, or for neutral penetrants, the solution--diffusion theory~\cite{wijmans1995solution} provides a very simple and fundamental relation, according to which the permeability $P$ is the product of the diffusion coefficient and partitioning, $P=KD$.
Based on this, the permeability of a molecule can be seen as an outcome of a complex and competing interplay between diffusion and solubility.
As it turns out, $K$ and $D$ are quite often anti-correlated for different morphologies of a polymer or for different penetrants in a given polymer. That is to say, a penetrant that tends to diffuse fast through a given material typically sorbs weakly in there. On the contrary, a penetrant that tends to sorb well generally diffuses slowly. The net effect of this trade-off is that the product $P=KD$ becomes less sensitive to various parameters than either the diffusivity or solubility~\cite{palasis1992permeability, ban2011molecular, kucukpinar2003molecular,  novitski2015determination, kusoglu2017new, kim2019tuning, kim2020tuning, kanduc2021shape}. 
Because of the cancellations, the permeability depends on tiny details of the polymer matrix and the penetrant.

The fact that slight differences in penetrants can result in substantial differences in their permeabilities gives rise to selectivity---the ability of a membrane to selectively pass one type of ions but not others~\cite{kanduc2021shape}.
In materials science, tailoring selective transport of ions and molecules through polymer membranes and other porous materials is of utmost importance for applications ranging from water desalination and filtration to drug delivery~\cite{park2017maximizing}.

The universally low ionic permeability compared with neutral molecules, such as water, is exploited in state-of-the-art desalination membranes (e.g., polyamide), which offer great water--salt selectivity. However, their ability to discriminate between ions is fairly limited~\cite{zhou2020intrapore, epsztein2020towards}. 
Yet, the demand for ion--ion selectivity is rapidly growing, for example, in the recovery of valuable ions from seawater (e.g., lithium and uranium) to mitigate resource shortages or in the development of new battery systems. 
In lithium-ion batteries, solid amorphous polymers [most notably poly(ethylene) oxide (PEO)] are regarded as attractive candidates to replace today’s liquid organic electrolytes~\cite{molinari2018effect, widstrom2021water}.
Improvements in the selective transport of lithium ions over other ions often rely on plasticizing the polymer network by adding various materials.
It has been shown, for instance, that introducing water into the PEO matrix enormously improves the conductance of Li$^+$ compared to other anions. Simulations revealed that these exceptional transport properties arise from strong lithium solvation and diffusion in percolated water nanodomains~\cite{widstrom2021water}.

Of particular interest in selective ionic transport is also the transport of protons.
In an aqueous environment, a proton manifests as the hydronium ion (H$_3$O$^+$) and can diffuse via two complementary mechanisms. The first mechanism is a classical center-of-mass motion of the hydronium ion, termed the vehicular mechanism. In the second, termed the Grotthuss mechanism, the excess proton hops from the hydronium ion across the hydrogen bond network of water, which is possible because of low barriers in the proton energy landscape~\cite{kusoglu2017new, fischer2018correlated}. 
This hopping ability gives protons in water an anomalously large diffusion coefficient, which is up to 7 times that of similarly sized cations~\cite{fischer2018correlated,peng2018transport, mabuchi2018relationship, okuwaki2018theoretical, vishnyakov2018coarse, huo2019molecular}. 
A group of synthetic polymers that are nowadays maybe the most explored in terms of perm-selective proton conductivity belongs to perfluorinated sulfonic-acid (PFSA) ionomers, such as Nafion, mentioned above.
Narrow water channels, many times even single-file water wires, in a PFSA polymer allow the diffusion of protons, but to a much lesser extent the diffusion of other ions. This property makes PFSA polymers widely used as perm-selective conductive membranes in various electrochemical technologies, including fuel cells, and as diffusion protection barriers against various toxic and waste chemicals~\cite{kusoglu2017new}.

In the end, we glance at biological ion channels, which provide incredible ion--ion selectivity.
For instance, potassium channels transport K$^+$ ions 10,000 times faster than Na$^+$ ions through the cell membrane~\cite{epsztein2020towards}. Furthermore, transmembrane proteins such as cytochrome {\sl c} oxidase, photosystem II, channelrhodopsin, and bacteriorhodopsin extremely selectively conduct protons through internal single-file water wires. 
These inspirational examples from nature can offer guidelines for ultra-high ion--ion selectivity in the fabrication of modern synthetic membranes. 
Despite the progress, the selectivity of synthetic membranes is often modest or limited to a particular property (e.g., divalent cations)~\cite{epsztein2020towards}.
Without a doubt, to fabricate novel polymer materials with high ion--ion selectivity, there is a crucial need to obtain a better understanding of the mechanisms for diffusion and solvation.

\section{Conclusions}
\label{sec:conclusions}

The transport of ions is a ubiquitous process in a vast range of different materials. The detailed knowledge of how ions diffuse and solvate in these materials is not only key to understand nature but also to devise a desired property of synthetic materials.
Weakly hydrated polymer membranes are more challenging to understand than highly hydrated, swollen polymers, owing to a much more intricate interplay between molecular interactions in the former and consequently a higher sensitivity to details on molecular structure. Yet, 
on the flip side, weakly hydrated systems offer more possibilities for a fine-tuned selective transport of ions.

In weakly hydrated polymers, water organizes into isolated nanoclusters---droplets of a high dielectric constant in an otherwise hydrophobic, low-dielectric surrounding of the polymer. Already a simple continuum picture provides a clue that ions tend to strongly solvate in these water nanoclusters, which is supported by simulations. Detailed simulations and resulting analysis also reveal that the interface potential of water clusters, being unimportant for most cases with simple ions, can become critical for molecular ions that are less hydrated because of a smeared charge.
Thanks to modern simulation approaches, the field of polymer science has started unraveling the fine details of the diffusion of neutral molecules through dense polymers. However, our understanding of ion diffusion is still very limited. The diffusion of ions involves several intertwined molecular mechanisms, which increase the complexity of the problem.
General conclusions reached so far are that ions diffuse slower than similarly-sized neutral molecules and that ion-specific effects turn out to be crucial.

The diffusivity and solvation of penetrants generally have opposite trends for different morphologies of a polymer or for different penetrants. Consequently, the resulting permeability, which is the product of the two, encounters enormous cancellation effects and depends on tiny molecular details.
Obviously, the number of chemical ways to synthesize a polymer membrane (e.g., with different combinations of copolymerization) is essentially unlimited. Synthesizing or modeling diverse polymer systems to optimize a desired property or functionality is thus out of reach. The fundamental understanding of the underlying phenomena is thus a prerequisite to attain the ambitious goal.

\section{Acknowledgments}
This project has received funding from the European Research Council (ERC) under the European Union’s Horizon 2020 research and innovation program (Grant Agreement No.\ 646659-NANOREACTOR). 
M.K.\ acknowledges the financial support from the Slovenian Research Agency (contracts P1-0055 and J1-1701). W.K.K. acknowledges the support by a KIAS Individual Grant (CG076001) at Korea Institute for Advanced Study.


\bibliographystyle{elsarticle-num}
\bibliography{COCISreview}

\end{document}